\documentstyle[12pt,fleqn]{article}
\input tcilatex
\QQQ{Language}{
American English
}

\begin{document}

\setlength{\textwidth}{600mm} \setlength{\textheight}{240mm} \voffset=-25mm 
\baselineskip=20pt plus 2pt

\begin{center}
{\large {\bf The Energy of the Gamma Metric in the M{\o }ller Prescription}}%
\\\vspace{5mm} \vspace{5mm} I-Ching Yang$^{\dagger }$ \footnote{%
E-mail:icyang@dirac.phys.ncku.edu.tw} and Irina Radinschi$^{\ddagger }$ 
\footnote{%
E-mail:iradinsc@phys.tuiasi.ro} \\

$^{\dagger}$Department of Natural Science Education, \\National Taitung
Teachers College, \\Taitung, Taiwan 950, Republic of China \\and \\$%
^{\ddagger}$Department of Physics, ``Gh. Asachi" Technical University, \\%
Iasi, 6600, Romania
\end{center}

\vspace{5mm}

\begin{center}
{\bf ABSTRACT}

We obtain the energy distribution of the gamma metric using the
energy-momentum complex of M{\o}ller. The result is the same as

obtained by Virbhadra in the Weinberg prescription.
\end{center}

\vspace{2mm} \noindent
 \newpage

\section{INTRODUCTION}

Energy-momentum is regarded as the most fundamental conserved quantity in
physics, and associated with a symmetry of space-time geometry. According to
Noether's theorem and translations invariance, one could define a conserved
energy-momentum $T^{\mu \nu }$ as a consequence of its satisfying the
differential conservation law $\partial _\nu T^{\mu \nu }=0$. However, in a
curve space-time where the gravitational field is presented, the
differential conservation law becomes 
\begin{equation}
\nabla _\nu T^{\mu \nu }=\frac 1{\sqrt{-g}}\frac \partial {\partial x^\nu }(%
\sqrt{-g}T^{\mu \nu })-\frac 12g^{\nu \rho }\frac{\partial g^{\nu \rho }}{%
\partial x^\lambda }T^{\mu \lambda }=0,
\end{equation}
and generally does not lead to any conserved quantity. Early energy-momentum
investigations attempted to determine the conserved energy-momentum for the
gravitational field and the matter located in it, and led to energy-momentum
complex 
\begin{equation}
\Theta ^{\mu \nu }=\sqrt{-g}(T^{\mu \nu }+t^{\mu \nu }),
\end{equation}
which satisfies the differential conservation equation $\partial _\nu \Theta
^{\mu \nu }=0$. Here, $T^{\mu \nu }$ is the energy-momentum tensor of matter
and $t^{\mu \nu }$ is regarded as the contribution of energy-momentum from
the gravitational field. There are various energy-momentum complexes,
including those of Einstein~\cite{1}, Tolman~\cite{2}, Papapetrou~\cite{3},
Bergmann~\cite{4}, Landau and Lifshitz~\cite{5}, M{\o }ller~\cite{6}, and
Weinberg~\cite{7}. On the other way, a different idea, quasilocal (i.e.,
associated with a closed 2-surface) was proposed. The Hamiltonian for a
finite region, 
\begin{equation}
H(N)=\int_\Sigma N^\mu \,{\cal H}_\mu +\oint_{S=\partial \Sigma }{\cal B}(N),
\end{equation}
generates the space-time displacement of a finite spacelike hypersurface $%
\Sigma $ along a vector field $N^\mu $. Noether's theorem guarantee that $%
{\cal H}_\mu $ is proportional to the filed equation. Consequently, the
value depends only on the boundary term ${\cal B}$, which gives the
quasilocal energy-momentum. Moreover, there are also a large number of
definitions of quasilocal mass~\cite{8,9}. In their recent article, Chang 
{\it et al.}~\cite{9} showed that every energy-momentum complex can be
associated with a particular Hamiltonian boundary term. So the
energy-momentum complexes may also be considered as quasilocal.

Though Penrose~\cite{10} points out that a quasilocal mass is conceptually
important. However, Bergqvist~\cite{11} studied several different
definitions of quasilocal masses for the Reissner- Nordstr\"{o}m and Kerr
space-times and came to the conclusion that not even two of these
definitions gave the same results. On the contrary, several energy-momentum
complexes have been showing a high degree of consistency in giving the same
energy distribution for a given space-time. Recently, Virbhadra and his
collaborators~\cite{12,13,14,15,16} have investigated that for a given
space-time (like as the Kerr- Newman, the Vaidya, the Einstein-Rosen, the
Bonnor-Vaidya and all Kerr-Schild class space-time) different
energy-momentum complexes (the Einstein, the Landau-Lifshitz, the
Papapetrou, the Tolman, The Weinberg, etc.) give the same energy
distribution. Moreover some interesting results~\cite{12,17,18,19,20} led to
the conclusion that in a given space-time (the Reissner-Nordstr\"{o}m, the
Kerr-Newman, the Garfinkle-Horowitz-Strominger, the de Sitter-Schwarzschild,
and the charged regular metric, etc.) the energy distribution according to
the energy-momentum complex of M{\o }ller is different from of Einstein. But
in some specific case~\cite{6,17} (the Schwarzschild, the
Janis-Newman-Winicour metric, etc.) there are the same. Recently, the energy
distribution in the Weinberg prescription obtained by Virbhadra~\cite{21}
using the gamma metric, is given as 
\begin{equation}
E=m\,\gamma .
\end{equation}
So, in this letter, we evaluate the energy distribution of the gamma metric
by using M{\o }ller energy-momentum complex, and compare with the result
obtained by Virbhadra with Weinberg energy-momentum complex.

\section{ENERGY\ IN\ THE\ M\O LLER\ PRESCRIPTION}

First, the well-known gamma metric~\cite{21,22}, a static and asymptotically
flat exact solution of Einstein vacuum equations, is given as 
\begin{equation}
ds^2=(1-\frac{2m}r)^\gamma dt^2-(1-\frac{2m}r)^{-\gamma }\left[ \left( \frac
\Delta \Sigma \right) ^{\gamma ^2-1}dr^2+\frac{\Delta ^{\gamma ^2}}{\Sigma
^{\gamma ^2-1}}d\theta ^2+\Delta sin^2\theta d\phi ^2\right] ,
\end{equation}
where 
\begin{eqnarray}
\Delta &=&r^2-2mr,  \nonumber \\
\Sigma &=&r^2-2mr+m^2sin^2\theta .
\end{eqnarray}
For $\left| \gamma \right| =1$ the metric is spherically symmetric and for $%
\left| \gamma \right| \neq 1$, it is axially symmetric. In the situation $%
\left| \gamma \right| =1$, the gamma metric reduces to the Schwarzschild
space-time. However, in another situation $\left| \gamma \right| \neq 1$,
the gamma metric gives the Schwarzschild space-time with negative mass, as
putting $m=-M(M>0)$ and carrying out a coordinate transformation $%
r\rightarrow R=r+2M$ one gets the Schwarzschild space-time with positive
mass.

Next, let us consider the M{\o }ller energy-momentum complex which is given
by~\cite{6} 
\begin{equation}
\Theta _\nu ^{\,\;\;\mu }=\frac 1{8\pi }\frac{\partial \chi _\nu ^{\;\;\mu
\sigma }}{\partial x^\sigma },
\end{equation}
where the M{\o }ller superpotential, 
\begin{equation}
\chi _\nu ^{\;\;\mu \sigma }=\sqrt{-g}\left( \frac{\partial g_{\nu \alpha }}{%
\partial x^\beta }-\frac{\partial g_{\nu \beta }}{\partial x^\alpha }\right)
g^{\mu \beta }g^{\sigma \alpha },
\end{equation}
are quantities antisymmetric in the indices $\mu $, $\sigma $. According to
the definition of the M{\o }ller energy-momentum complex, the energy
component is given as 
\begin{eqnarray}
E &=&\int \Theta _0^{\;\;0}dx^1dx^2dx^3  \nonumber \\
&=&\frac 1{8\pi }\int \frac{\partial \chi _0^{\;\;0k}}{\partial x^k}%
dx^1dx^2dx^3,
\end{eqnarray}
where the Latin index takes values from 1 to 3. However, in the case, the
only nonvanishing component of M{\o }ller's superpotential is 
\begin{equation}
\chi _0^{\;\;01}=2\,m\,\gamma \,sin\theta .
\end{equation}
Applying the Gauss theorem to (9) and using (10), we evaluate the integral
over the surface of a sphere with radius $r$, and find the energy
distribution is 
\begin{equation}
E=m\,\gamma .
\end{equation}
It is the same result as obtained by Virbhadra in the Weinberg prescription.

\section{DISCUSSION}

It is well-known that the subject of the energy-momentum localization is
associated with much debate. In contradiction with Misner {\it et al.}[23],
Cooperstock and Sarracino [24] gave their viewpoint that if the energy
localization is meaningful for spherical system it is, also, meaningful for
all systems. Also, Cooperstock [25] gave his opinion that the energy and
momentum are confined to the regions of non-vanishing energy-momentum tensor
of the matter and all non-gravitational fields. Bondi [26] sustained that a
nonlocalizable form of energy is not admissible in relativity so its
location can in principle be found.

We calculate the energy distribution of the gamma metric using the
energy-momentum complex of M\o ller. The energy depends on the mass $m$.
Thus, we get the same result as Virbhadra [21] obtained using the
energy-momentum complex of Weinberg. This result sustains the opinion that
different energy-momentum complexes could give the same expression for the
energy distribution in a given space-time. As we noted, for some given
space-times [17] the energy distribution according with the energy-momentum
complex of M\o ller is the same as those calculated in the Einstein
prescription. Our results sustain the conclusion of Lessner [27] that the
M\o ller energy-momentum complex is an important concept of energy and
momentum in general relativity. Also, the M\o ller energy-momentum complex
allows to make the calculations in any coordinate system.

\end{document}